\documentclass[aps,prb,twocolumn]{revtex4-1}
\usepackage{graphicx}
\usepackage{pstricks}
\usepackage{amsmath}
\usepackage{dcolumn}
\usepackage{nicefrac}

\begin{document}
\title{Reduction of magnetic interlayer coupling in barlowite through isoelectronic substitution}

\author{Daniel Guterding}
\email{guterding@itp.uni-frankfurt.de}
\affiliation{Institut f\"ur Theoretische Physik, Goethe-Universit\"at Frankfurt, 
Max-von-Laue-Stra{\ss}e 1, 60438 Frankfurt am Main, Germany}

\author{Roser Valent\'i}
\affiliation{Institut f\"ur Theoretische Physik, Goethe-Universit\"at Frankfurt,
Max-von-Laue-Stra{\ss}e 1, 60438 Frankfurt am Main, Germany}

\author{Harald O. Jeschke}
\affiliation{Institut f\"ur Theoretische Physik, Goethe-Universit\"at Frankfurt, 
Max-von-Laue-Stra{\ss}e 1, 60438 Frankfurt am Main, Germany}

\begin{abstract}
  Materials with a perfect kagome lattice structure of magnetic ions
  are intensively sought for, because they may exhibit exotic ground
  states like the a quantum spin liquid phase. Barlowite is a natural
  mineral that features perfect kagome layers of copper ions. However,
  in barlowite there are also copper ions between the kagome layers,
  which mediate strong interkagome couplings and lead to an ordered
  ground state. Using {\it ab initio} density functional theory
  calculations we investigate whether selective isoelectronic
  substitution of the interlayer copper ions is feasible. After
  identifying several promising candidates for substitution we
  calculate the magnetic exchange couplings based on crystal
  structures predicted from first-principles calculations. We find
  that isoelectronic substitution with nonmagnetic ions significantly
  reduces the interkagome exchange coupling. As a consequence,
  interlayer-substituted barlowite can be described by a simple
  two-parameter Heisenberg Hamiltonian, for which a quantum spin
  liquid ground state has been predicted.
\end{abstract}


\maketitle

\section{Introduction}
The synthesis of herbertsmithite and similar frustrated
antiferromagnets with a spin-1/2 kagome lattice has generated in the
past intense research efforts because these materials possibly realize
a quantum spin liquid (QSL) phase~\cite{HerbertsmithiteSynthesis,
  Mendels2010, Mendels2011, HerbertsmithiteExchange, Colman2010,
  Li2013, Li2014, Han2012, MgHerbertsmithite, Yan2011, BauerAnyons,
  TopoDMRG, Mazin2014, Guterding2016, Norman2016, Iqbal2015}. However,
a problem in herbertsmithite and related materials is the intrinsic
disorder~\cite{HerbertsmithiteDisorder, AntisiteDisorder,
  AntisiteDisorder2}, which makes the interpretation of experimental
data more difficult and might even prevent the formation of a QSL.
Therefore, chemical alternatives to herbertsmithite with slightly
different bonding environment are intensively sought
for~\cite{Norman2016}.

Recently, the geometrically perfect spin-1/2 kagome antiferromagnet
barlowite [Cu$_4$(OH)$_6$FBr] was discovered~\cite{Elliott2014,
  Han2014, BarlowiteTheoryExperiment, Liu2015}. It is closely related
to herbertsmithite, but shows a different stacking pattern of the
kagome layers and features magnetic ions on interplane sites, which
couple the kagome planes magnetically and lead to magnetic ordering at
low temperatures~\cite{BarlowiteTheoryExperiment}. Selective doping of
non-magnetic ions into those interplane sites could be a way to remove
these interlayer couplings and realize a quantum spin liquid.

Previous calculations showed that substitution with several divalent
cations should be possible~\cite{Liu2015}. These authors predicted
that magnesium and zinc can selectively substitute the interlayer
copper atoms, but did not make a statement about the resulting
structure of the magnetic exchange paths.  As the interlayer distance
in barlowite is relatively small, a finite interlayer exchange can
however be expected, even when the magnetic ions between the kagome
layers are removed.

Using {\it ab initio} density functional theory (DFT) calculations we
investigate the energetics of selectively substituting the copper ions
between kagome layers in the barlowite system. In contrast to previous
speculations~\cite{Han2014} and in agreement with recent
calculations~\cite{Liu2015} we find that isoelectronic doping at
interlayer position becomes energetically less favorable with
increasing ionic radius of the substituent. Hence, we identify ions
with radius similar to Cu$^{2+}$ as the most promising candidates for
substitution, such as magnesium or zinc. In this respect we agree with
the calculations by Liu {\it et al.}~\cite{Liu2015}. We however find
that most other ions lead to monoclinically distorted crystal
structures with no perfect kagome arrangement of copper ions.

Based on the crystal structures predicted for those substitutions with
perfect kagome lattice we estimate the magnetic exchange couplings in
the substituted barlowite system. We find that the corresponding
Heisenberg Hamiltonian takes a very simple form with a parameter $J_1$
describing the nearest-neighbor kagome coupling and a parameter $J_2$
describing the inter-layer exchange energy. We compare our findings to
recent model calculations, which suggested a QSL for certain ratios of
$J_2 / J_1$~\cite{Goetze2015}.

\section{Methods}
Experimental and hypothetical crystal structures were fully relaxed
using the projector augmented wave (PAW) method~\cite{PAWmethod}
implemented in {\sc GPAW}~\cite{GPAWmethod} with a plane-wave cutoff
of $1000~\mathrm{eV}$ and the GGA exchange-correlation
functional~\cite{PerdewBurkeErnzerhof}. We optimized all crystal
structures using $4^3$ $k$-points until forces were below $10\,
\mathrm{meV/\AA}$.

We started from the experimental crystal structure of
barlowite~\cite{Elliott2014} and substituted the interlayer copper
site (Cu$^{2+}$) by
$A$=Mg$^{2+}$,~Ca$^{2+}$,~Sr$^{2+}$,~Zn$^{2+}$,~Cd$^{2+}$,~Hg$^{2+}$,~Sn$^{2+}$,
~Pb$^{2+}$, forming $A$-substituted barlowite,
$A$Cu$_3$(OH)$_6$FBr. From this relaxation we obtained the desirable
crystal structure with perfect kagome lattice of substituted barlowite
in space group $P\,6_3/mmc$, containing two formula units. While most
of the cations considered here are almost certainly divalent, tin and
lead are more flexible with respect to their oxidation
state. Therefore, it has to be verified that these cations indeed
assume a divalent state when substituted into barlowite.

\begin{table}[tb]
\caption{Calculated isoelectronic doping energies for barlowite. Positive values 
of the doping energy $E_\text{dop}$ indicate that the kagome lattice will be 
distorted upon doping. On the interlayer site, barlowite prefers to incorporate 
ions with smaller radius than Cu$^{2+}$ (72~pm). Ionic radii in coordination 
number 6 are taken from Refs.~\onlinecite{IonicRadii,Sn3IonicRadius}.}
\begin{ruledtabular}
\begin{tabular}{rrrrrrrr}
$A=$ & Mg$^{2+}$ & Ca$^{2+}$ & Sr$^{2+}$ & Zn$^{2+}$ & Cd$^{2+}$ & Hg$^{2+}$ & Sn$^{3+}$ \\ \hline
$E_\text{dop}$ (eV) & -0.994 & -0.012 & 0.349 & -0.367 & 0.183 & 1.282 & -0.196 \\
$r_\text{ion}$ (pm) & 72 & 100 & 118 & 74 & 95 & 102 & 81 \\
\end{tabular}
\end{ruledtabular}
\label{tab:energies}
\end{table}

\begin{figure}[tb]
\includegraphics[width=\linewidth]{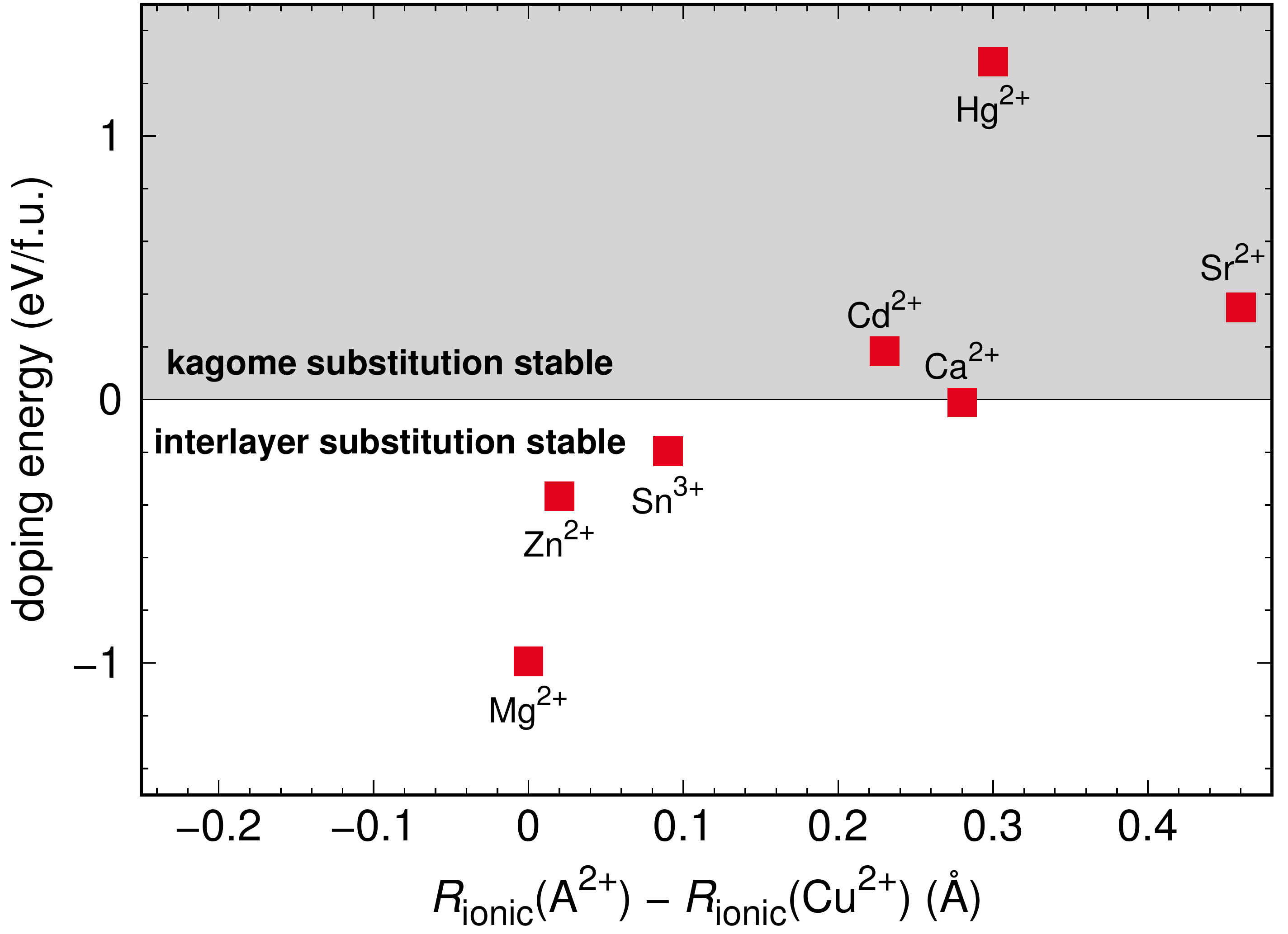}
\caption{(Color online) Calculated isoelectronic doping energies for
  barlowite.  All data points above the zero line indicate that the
  kagome lattice will be distorted upon doping. On the interlayer
  site, barlowite prefers to incorporate ions with smaller radius than
  Cu$^{2+}$ (72~pm). Ionic radii in coordination number 6 are taken
  from Refs.~\onlinecite{IonicRadii,Sn3IonicRadius}.}
\label{fig:dopingenergy}
\end{figure}

From these structures we then constructed related {\it undesirable}
defect structures, where all substituent atoms sit in the kagome
plane. After reducing the symmetry to $P\,1$, the unit cell contains three
copper atoms for each of two different kagome layers stacked in the
$z$-direction. We chose to place one dopant atom in each kagome layer
assuming that clustering is not favored. This reduces the number of
possible defect configurations to two. Either dopant atoms are stacked
in the $z$-direction or placed as far apart as possible, with
positioning in the $xy$-plane alternating between the two kagome
layers. We decided for the latter case. After full relaxation of these
structures with identical chemical composition $A$Cu$_3$(OH)$_6$FBr,
total energies can be compared directly on the DFT level to find
whether the substituent ions prefer to occupy the interlayer site or
occupy a site in the kagome layer, consequently destroying the perfect
kagome lattice.

Total energies and magnetic exchange interactions were evaluated based
on the fully relaxed structures using {\it ab initio} DFT calculations
within an all-electron full-potential local orbital
(\textsc{FPLO})~\cite{FPLOmethod} basis. For the exchange-correlation
functional we employed the generalized gradient approximation
(GGA)~\cite{PerdewBurkeErnzerhof}, as well as
GGA+U~\cite{Liechtenstein95} functionals. For the double counting
correction we used the fully localized limit. The Hubbard repulsion on
the Cu 3$d$ orbitals was set to $U=6~\mathrm{eV}$ and Hund's rule
coupling to $J_H = 1~\mathrm{eV}$.  This choice of parameters was
verified in Ref.~\onlinecite{BarlowiteTheoryExperiment} by comparing
experimental and simulated susceptibility data. For the calculation of
magnetic exchange couplings we used a $\sqrt{2}\times \sqrt{2}\times
1$ supercell containing four formula units, {\it i.e.} twelve
inequivalent Cu sites.  This allows for 171 unique out of 4096 total
spin configurations. The couplings we calculated are based on fits to
25 different spin configurations. For details of the fitting procedure,
see Appendix~\ref{app:heisenbergdetails}. As the Cu$^{2+}$ moments are
precisely 1~$\mu_{\rm B}$ and all spin configurations lead to gaps of
1.5~eV for $U=6$~eV and of 2.2~eV for $U=8$~eV, fits are very accurate
and the resulting statistical error bars tiny.  Total energies and
Heisenberg models were extracted from calculations converged using
$8^3$ and $6^3$ $k$-point grids respectively.  Although we found that
a sizeable Dzyaloshinskii-Moriya interaction is necessary to explain
the canted antiferromagnetic state in the original
barlowite~\cite{BarlowiteTheoryExperiment}, we do not investigate 
this parameter here, since it requires calculations beyond the scope
of the present study. In studies of the quantum Heisenberg
model with Dzyaloshinskii-Moriya interaction, it has been shown that
the QSL state we are interested in here is stable with substantial DM
values of up to $D \sim 0.1J$~\cite{Mendels2011,Messio2010}.

\section{Results and Discussion}
\begin{table}[t!]
  \caption{Predicted structural parameters for Mg-barlowite
  [MgCu$_3$(OH)$_6$FBr] ($P\,6_3/mmc$ space group,
    $a=6.79899$~{\AA}, $c=9.34513$~{\AA}, $Z=2$). }
\label{tab:structmg}
\begin{tabular}[c]{lllllll}
\hline\hline
Atom & Site & $x$ & $y$ & $z$ \\ \hline
Cu&$6g$&  $\nicefrac{1}{2}$     & 0& 0\\
Mg&$2c$&  $\nicefrac{1}{3}$ & $\nicefrac{2}{3}$& $\nicefrac{1}{4}$\\
O    &$12k$& 0.20346 &  0.40692 &  0.09419 \\
H    &$12k$&  0.12545 &  0.25090 &  0.13412 \\
Br   &$2d$&  $\nicefrac{1}{3}$ & $\nicefrac{2}{3}$& $\nicefrac{3}{4}$ \\
F    &$2b$&  0&0& $\nicefrac{1}{4}$\\
\hline\hline
\end{tabular}
\end{table}
\begin{table}[t!]
  \caption{Predicted structural parameters for Zn-barlowite
  [ZnCu$_3$(OH)$_6$FBr] ($P\,6_3/mmc$ space group,
    $a=6.81439$~{\AA}, $c=9.39923$~{\AA}, $Z=2$). }
\label{tab:structzn}
\begin{tabular}[c]{lllllll}
\hline\hline
Atom & Site & $x$ & $y$ & $z$ \\ \hline
Cu&$6g$&  $\nicefrac{1}{2}$     & 0& 0\\
Zn&$2c$&  $\nicefrac{1}{3}$ & $\nicefrac{2}{3}$& $\nicefrac{1}{4}$\\
O    &$12k$& 0.20333  & 0.40666  & 0.09228  \\
H    &$12k$&  0.12645  & 0.25290  & 0.13526  \\
Br   &$2d$&  $\nicefrac{1}{3}$ & $\nicefrac{2}{3}$& $\nicefrac{3}{4}$ \\
F    &$2b$&  0&0& $\nicefrac{1}{4}$\\
\hline\hline
\end{tabular}
\end{table}

\begin{table*}
\caption{Calculated exchange couplings for the theoretically predicted 
Mg-barlowite, MgCu$_3$(OH)$_6$FBr, and Zn-barlowite, ZnCu$_3$(OH)$_6$FBr. A 
GGA+U functional with  $J_H=1$~eV and the 
two listed $U$ values was used in combination with fully localized limit double 
counting correction~\cite{Anisimov93, Dudarev98}. Antiferromagnetic couplings 
are positive, while ferromagnetic couplings are negative. The statistical errors 
of the exchange parameters from the fitting procedure are given in brackets.}
\label{tab:couplings}
\begin{ruledtabular}
\begin{tabular}{lrrrrrrrr|r}
$A$=&$U$\,(eV) & $J_1$\,(K) & $J_2$\,(K) & $J_3$\,(K) & $J_4$\,(K) & $J_5$\,(K) & $J_6$\,(K) & $J_7$\,(K) & $J_2 / J_1$ \\
Mg$^{2+}$&6.0&226(1)&13.4(6)&0.0(4)&1.1(4)&-1.0(4)&0.1(3)&-0.4(2)&0.06\\
&8.0&167(1)&10.0(4)&0.1(2)&0.4(3)&-0.8(2)&0.0(2)&-0.2(1)&0.06\\
Zn$^{2+}$&6.0&240(1)&15.4(5)&0.8(2)&0.8(3)&-1.0(2)&-0.7(2)&-0.7(2)&0.07\\
&8.0&179(1)&11.0(3)&0.5(2)&0.1(2)&-0.9(2)&-0.4(2)&-0.4(1)&0.07\\
\end{tabular}
\end{ruledtabular}
\end{table*}

\begin{figure}[tb]
\includegraphics[width=\linewidth]{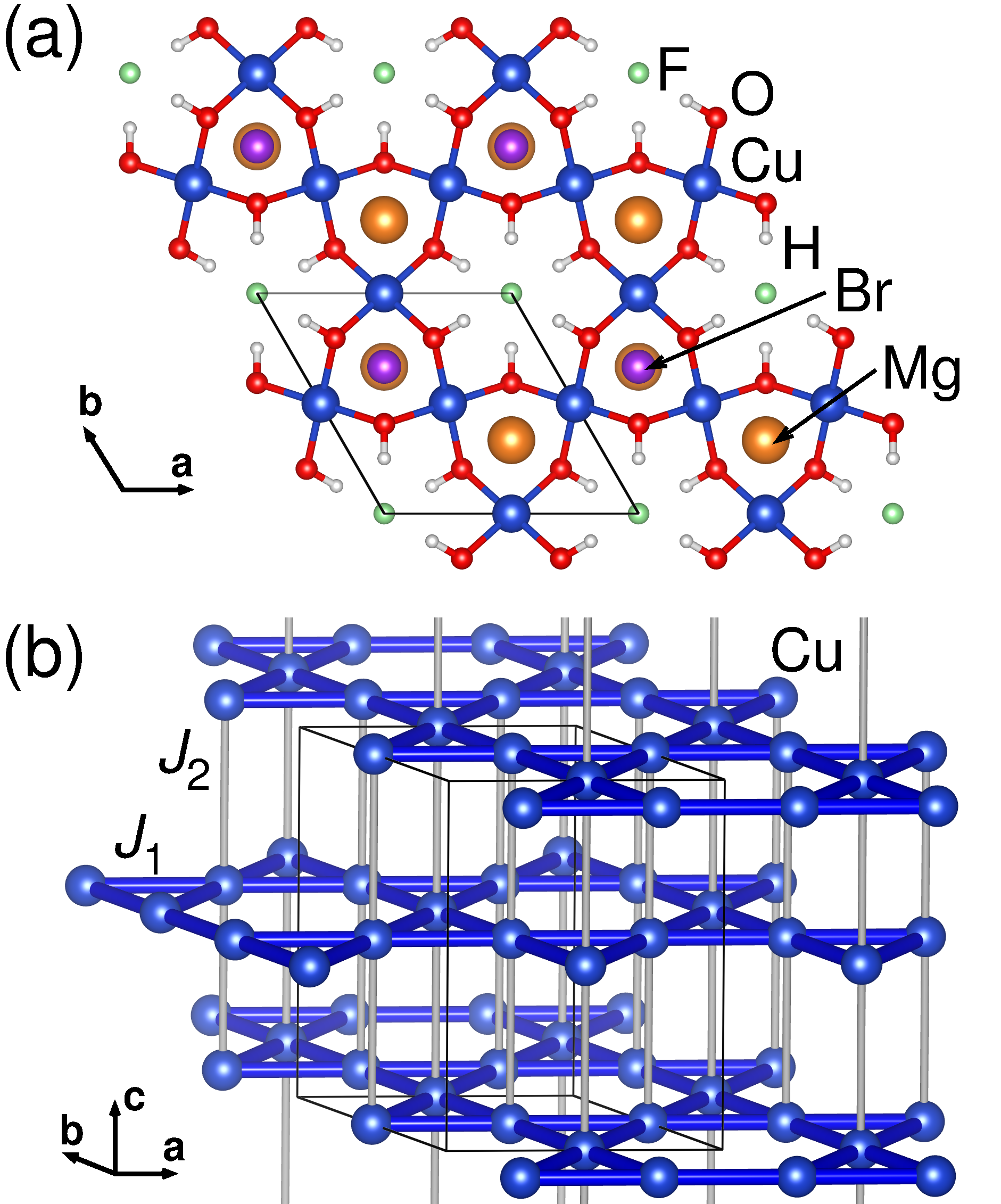}
\caption{(Color online) (a) Crystal structure and (b) magnetic exchange paths in Mg-barlowite, 
MgCu$_3$(OH)$_6$FBr.}
\label{fig:exchangepaths}
\end{figure}

The DFT calculated doping energies are given in
Table~\ref{tab:energies}. Our results show that substitution happens
selectively on the interlayer copper site, i.e. structures with
perfect kagome layer are obtained only for substitutions of the
interlayer Cu$^{2+}$ by magnesium, zinc and tin. For substitution with
calcium the energy difference between doping into the interlayer and
into the kagome plane is negligible. For strontium, cadmium, mercury
and lead we find that these substituents prefer to distort the kagome
plane. For lead not even a metastable structure with perfect kagome
layer could be obtained, therefore no doping energy has been
calculated. Surprisingly, we find that tin assumes a configuration
Sn$^{3+}$ when substituted into barlowite, which lowers the oxidation
state of the Cu$^{2+}$ ions by one third of an electron on average.
Consequently the magnetic moments in Sn-barlowite are very fragile and
we did not investigate the magnetic exchange couplings of this system
any further. As Sn-barlowite might still be an interesting material,
some details on its electronic structure are given in
Appendix~\ref{app:snbarlowitelectronic}.

In Fig.~\ref{fig:dopingenergy} we visualize the doping energy per
formula unit of $A$-substituted barlowite as a function of the ionic
radius in coordination number six, taken from
Refs.~\onlinecite{IonicRadii,Sn3IonicRadius}. The doping energy is
mainly controlled by the ionic radius of the substituents. The
chemical details of the substituents only play a minor role, evidenced
by the larger absolute doping energy of magnesium compared to zinc or
the negative doping energy of calcium compared to a positive doping
energy of cadmium, although the latter has a slightly smaller ionic
radius.

The doping energies we calculated reproduce the overall trends
reported in Ref.~\onlinecite{Liu2015}. As we performed our
calculations directly in the stoichiometric limit, we however found a
competing monoclinically distorted crystal structure, which places
stronger restrictions on the feasibility of the doping process than
the supercell calculations in Ref.~\onlinecite{Liu2015}.

The predicted crystal structure for Mg-barlowite [MgCu$_3$(OH)$_6$FBr]
is given in Table~\ref{tab:structmg} and visualized in
Fig.~\ref{fig:exchangepaths}(a).  The variant with zinc, Zn-barlowite
[ZnCu$_3$(OH)$_6$FBr], has a very similar structure listed in
Table~\ref{tab:structzn}.

For these optimized crystal structures we calculated the magnetic
exchange couplings as described previously. The calculated exchange
parameters for Mg-barlowite and Zn-barlowite are listed in
Table~\ref{tab:couplings}. The coupling constants are sorted so that
the index increases with the Cu-Cu distance along the bond. Couplings
$J_1$, $J_4$ and $J_5$ connect to the first, second and third
neighbors within the kagome plane, while $J_2$, $J_3$, $J_6$ and $J_7$
are interlayer couplings with a Cu-Cu distance of up to 8.3~\AA. We
find that only the exchange parameters $J_1$ and $J_2$ are
relevant. The corresponding paths are visualized in
Fig.~\ref{fig:exchangepaths}(b). The resulting spin-\nicefrac{1}{2}
Heisenberg Hamiltonian therefore contains antiferromagnetic couplings
$J_1$ along the nearest-neighbor bonds in the kagome plane, indicated
by symbol $\langle i,j \rangle$, and antiferromagnetic couplings $J_2$
along bonds perpendicular to the kagome layer, indicated by symbol
$[i,j]$:
$$
H = J_1 \sum \limits_{ \langle i,j \rangle} \vec S_i \cdot \vec S_j
+ J_2 \sum \limits_{[i,j]} \vec S_i \cdot \vec S_j
$$

In the sums, each bond is only counted once. 
The exchange parameters in Zn-barlowite are somewhat larger than in
Mg-barlowite. The larger $J_1$ in Zn-barlowite may be a consequence of
the Cu-O-Cu bond angles of $117.8^\circ$ (Zn-barlowite) versus
$117.2^\circ$ (Mg-barlowite).  The antiferromagnetic contribution to
the superexchange is enhanced when the bond angle gets closer to
$180^\circ$. The same evolution of exchange interactions with bond
angle is observed in the similar compounds herbertsmithite,
kapellasite and haydeeite~\cite{Iqbal2015},
(Zn,Mg,Cd)Cu$_3$(OH)$_6$Cl$_2$. The slightly larger $J_2$ in
Zn-barlowite could be related to the larger ionic radius of Zn$^{2+}$
compared to Mg$^{2+}$, which could enhance the exchange
interaction. Taking into account the statistical errors listed in
Table~\ref{tab:couplings}, the difference is however very small.

For the original barlowite system we estimated Heisenberg couplings
$J_{b1} = 177~\mathrm{K}$ for next-neighbor exchange within the kagome 
plane and $J_{b2} = -205~\mathrm{K}$ between spins within
the kagome plane and interlayer copper spins~\cite{BarlowiteTheoryExperiment}.
The strong interlayer coupling is clearly removed upon substitution
and the intralayer coupling retains the same order of magnitude.
With $U = 6~\mathrm{eV}$ we estimate a ratio of $J_2 / J_1 = 0.06$ for
Mg-barlowite and $J_2 / J_1 = 0.07$ for Zn-barlowite. A QSL has
recently been predicted~\cite{Goetze2015} for this $J_1$-$J_2$ model
in the range $| J_2 / J_1 | < 0.15$. Thus, we predict the two
barlowite variants Mg-barlowite and Zn-barlowite to be not only stable
new materials, but also very promising spin liquid candidates.

\section{Summary}
In summary, we performed {\it ab initio} density functional theory calculations 
for isoelectronically substituted barlowite. We showed that only magnesium and 
zinc can be expected to selectively substitute the interlayer copper atoms in 
barlowite. Based on the {\it ab initio} relaxed structures for Mg-barlowite and 
Zn-barlowite we predicted that these systems are accurately described by a 
two-parameter Heisenberg model including the nearest-neighbor kagome exchange 
and the exchange perpendicular to the kagome layers. According to our 
calculations both Mg-barlowite and Zn-barlowite lie in the region of model 
parameters for which a quantum spin liquid ground state has recently been 
predicted. Therefore, we expect that the synthesis of either of these 
substituted barlowite systems will provide an alternative to the established 
herbertsmithite system.

\begin{acknowledgments}
The authors thank Christian Klein and Cornelius Krellner for fruitful discussions. 
This work was supported by the German Research Foundation (Deutsche 
Forschungsgemeinschaft) through grant SFB/TR 49.
\end{acknowledgments}

\appendix
\section{Electronic Structure of Sn-barlowite in comparison to Mg-barlowite}
\label{app:snbarlowitelectronic}
The predicted crystal structure for Sn-barlowite is listed in 
Table~\ref{tab:structsn}. Compared to the compounds with the smaller magnesium 
and tin ions (see Tables~\ref{tab:structmg} and \ref{tab:structzn}) the $a$-axis 
is somewhat contracted, while the $c$-axis is significantly larger. The 
difference in the $c$-axes of Mg-barlowite and Zn-barlowite can be directly 
explained from the different size of Mg$^{2+}$ and Zn$^{2+}$ ions. In 
Sn-barlowite the c-axis expansion is much larger than the difference in ionic 
radius. Therefore, the expansion mechanism must be more complex.

\begin{table}[t]
  \caption{Predicted structural parameters for Sn-barlowite
  [SnCu$_3$(OH)$_6$FBr] ($P\,6_3/mmc$ space group,
    $a=6.73194$~{\AA}, $c=9.99153$~{\AA}, $Z=2$). }
\label{tab:structsn}
\begin{tabular}[c]{lllllll}
\hline\hline
Atom & Site & $x$ & $y$ & $z$ \\ \hline
Cu&$6g$&  $\nicefrac{1}{2}$     & 0& 0\\
Sn&$2c$&  $\nicefrac{1}{3}$ & $\nicefrac{2}{3}$& $\nicefrac{1}{4}$\\
O    &$12k$& 0.20061 & 0.40122  & 0.10735 \\
H    &$12k$&  0.12368 & 0.24736 &  0.14979 \\
Br   &$2d$&  $\nicefrac{1}{3}$ & $\nicefrac{2}{3}$& $\nicefrac{3}{4}$ \\
F    &$2b$&  0&0& $\nicefrac{1}{4}$\\
\hline\hline
\end{tabular}
\end{table}

\begin{figure}[tb]
\includegraphics[width=\linewidth]{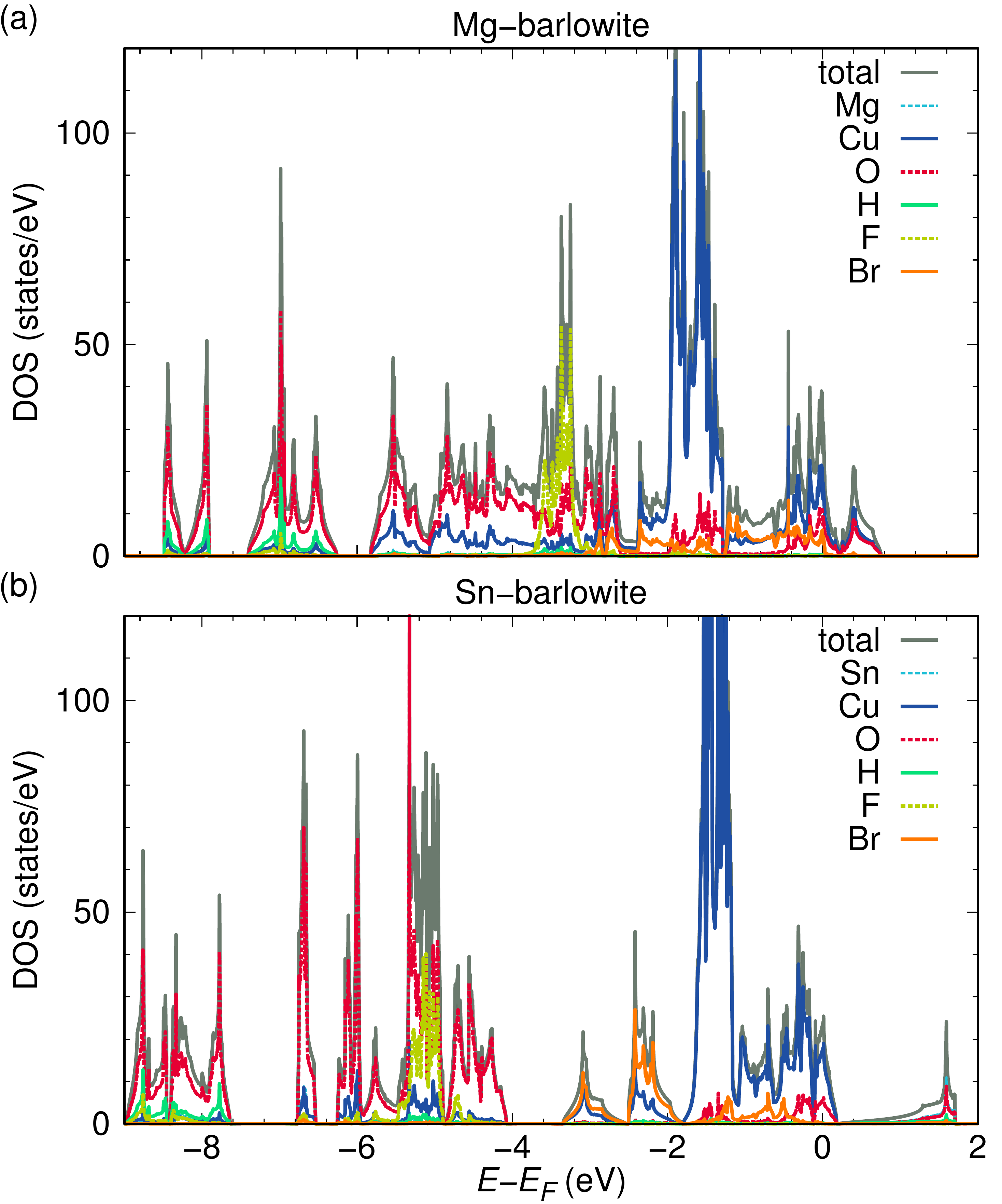}
\caption{(Color online) Calculated GGA density of states for (a) Mg-barlowite and (b) Sn-barlowite.}
\label{fig:dossn}
\end{figure}

\begin{figure}[tb]
\includegraphics[width=\linewidth]{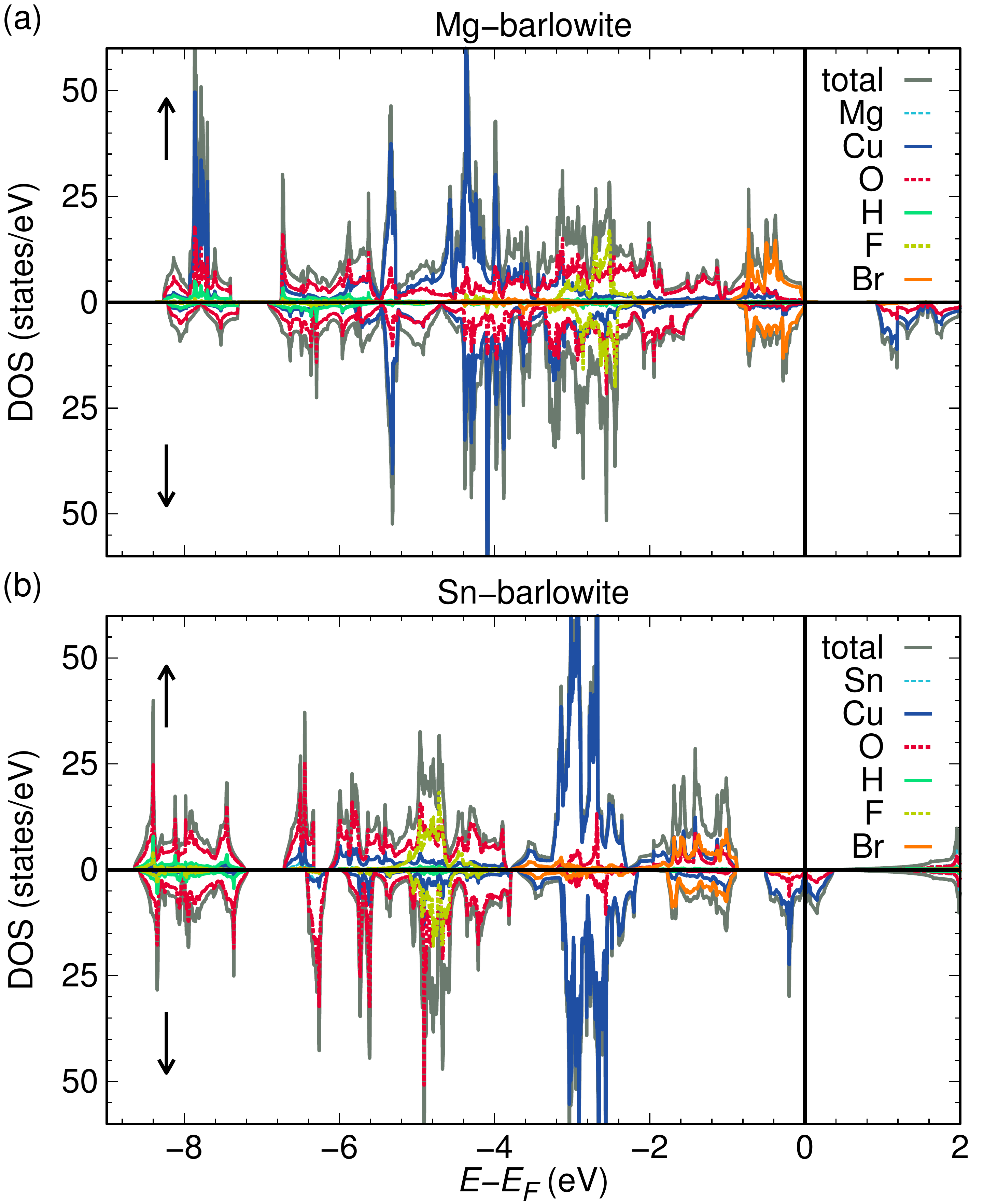}
\caption{(Color online) Calculated spin-resolved GGA+U ($U=6~\mathrm{eV}$, $J_H = 
1~\mathrm{eV}$) density of states for (a) Mg-barlowite and (b) Sn-barlowite in 
the ferromagnetic state. The top panel of each subplot shows the DOS of the 
minority spin channel, while the bottom panel shows the DOS of the majority spin 
channel.}
\label{fig:dossnggau}
\end{figure}

The calculated GGA density of states for Mg-barlowite and Sn-barlowite is shown 
in Fig.~\ref{fig:dossn}. The most prominent difference between the two materials 
is the distribution of the bromine and fluorine states, which are shifted to 
much lower energies in Sn-barlowite. Especially the hybridization between 
bromine and copper is very much reduced. 

Furthermore, integration of the copper density of states reveals that the 
occupation is one electron higher than in Mg-barlowite, while tin is in a 
trivalent state. We conclude that tin donates an additional electron into the 
copper kagome layer. This behavior can be rationalized based on the observation 
that doping onto the interlayer site becomes more stable with smaller ionic 
radius (see Fig.~\ref{fig:dopingenergy}). While Sn$^{2+}$ has an ionic radius of 
93~pm~\cite{Sn2IonicRadius}, Sn$^{3+}$ has an ionic radius of 81~pm. As a  
consequence Sn-barlowite can gain energy by reducing the ionic radius as tin 
becomes trivalent. 

The strong expansion of the $c$-axis can be explained by an elongation of the 
Cu-O bonds due to the additional charge on the copper atoms donated from tin. 
This reduction of Cu-O hybridization is directly visible in the density of 
states (see Fig.~\ref{fig:dossn}). This in turn might explain the large shift in 
bromine and fluorine energy levels compared to Mg-barlowite. Elongation of Cu-O 
bonds induces a shrinking of the F-H bonds, which leads to a strong lowering of 
the fluorine energy levels. Bromine on the other hand almost exclusively 
hybridizes with copper, and these Cu-Br bonds are significantly elongated in 
Sn-barlowite, probably again due to the additional charge available on the 
copper atoms. Therefore, the hybridization between bromine and copper is reduced 
and bromine states are less relevant in the vicinity of the Fermi level.

\begin{figure}[tb]
\includegraphics[width=\linewidth]{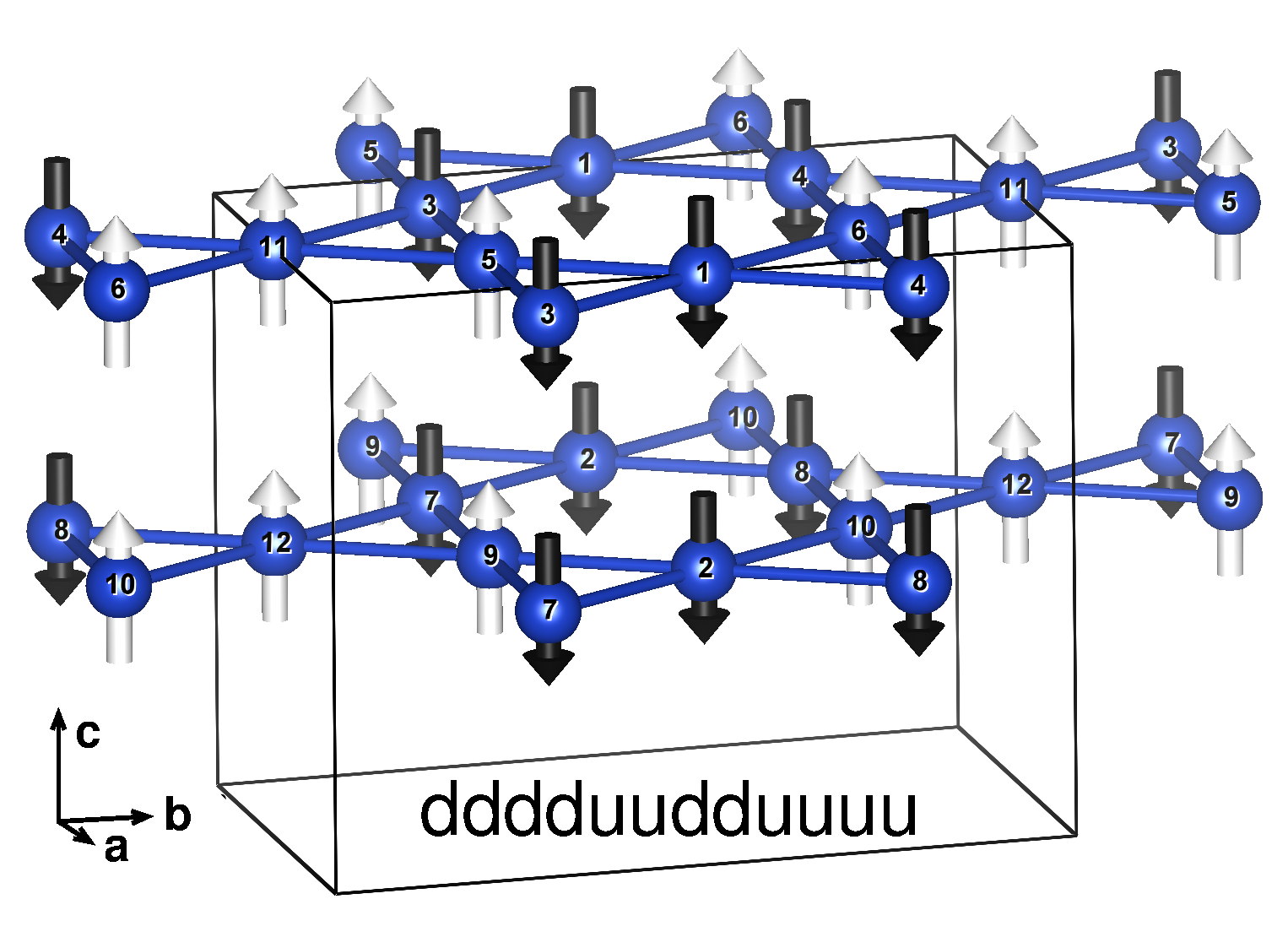}
\caption{(Color online) Example for the spin configurations used in
  the energy mapping. Only the Cu$^{2+}$ ion positions for a
  $\sqrt{2}\times \sqrt{2}\times 1$ supercell of Zn-barlowite are
  shown. The numbers on the twelve inequivalent Cu$^{2+}$ ions allow
  encoding of the spin configuration indicated with arrows as the
  string ``dddduudduuuu'' where d$\equiv \downarrow$ and u$\equiv
  \uparrow$.}
\label{fig:magneticconf}
\end{figure}

Additionally, we show in Fig.~\ref{fig:dossnggau} the spin-resolved
GGA+U density of states for ferromagnetic arrangement of copper
moments. In our calculations, Mg-barlowite clearly becomes insulating
once magnetism and interactions are considered, while Sn-barlowite
stays metallic due to to the additional electron in the copper kagome
layer, leading to $\nicefrac{4}{3}$ filling~\cite{Mazin2014}.

\section{Calculation of Heisenberg exchange couplings by mapping of DFT energies}
\label{app:heisenbergdetails}

\begin{figure}[tb]
\includegraphics[width=\linewidth]{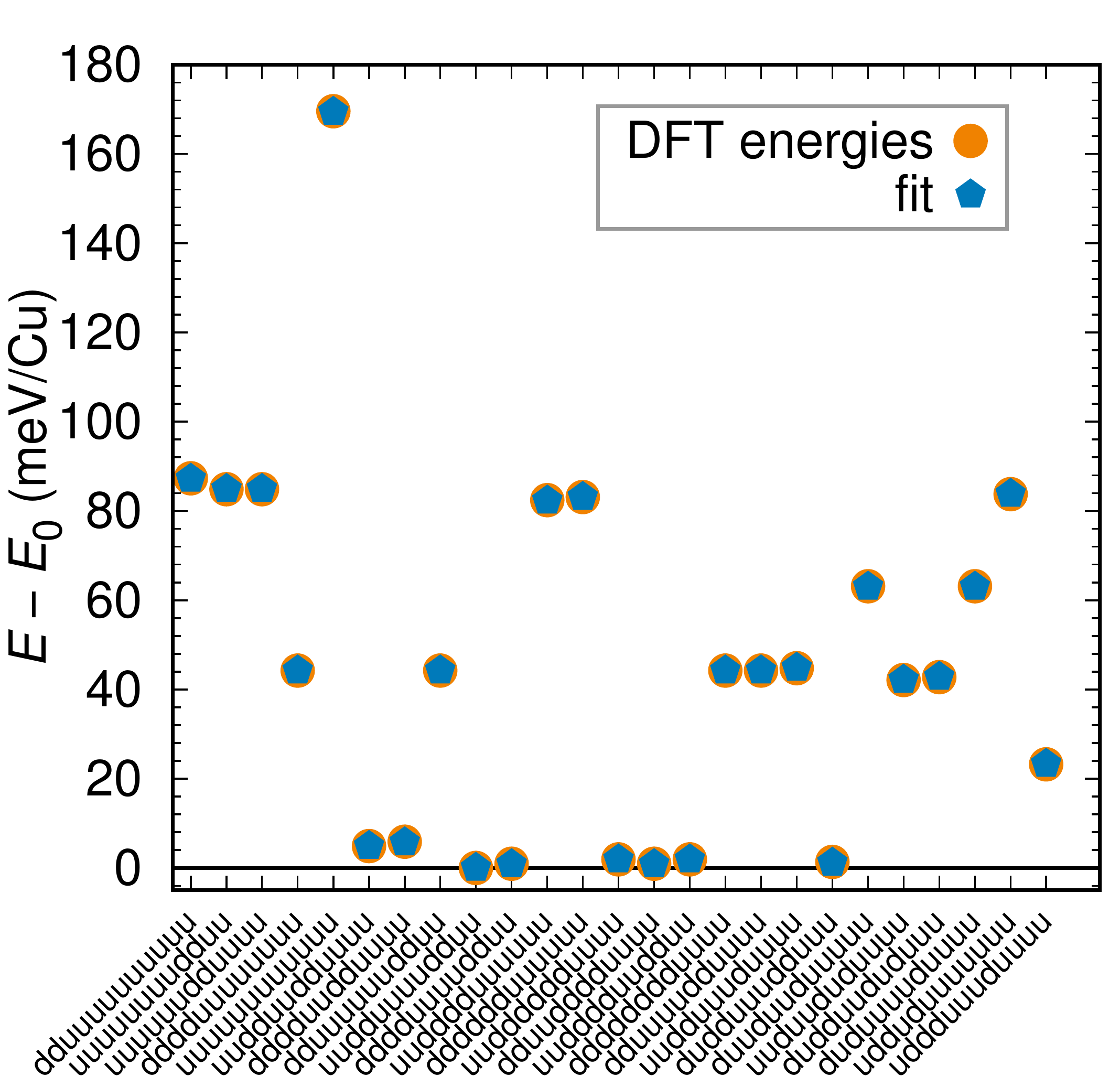}
\caption{(Color online) Comparison between calculated total energies
  per Cu$^{2+}$ ion for 25 different spin configurations in a
  $\sqrt{2}\times \sqrt{2}\times 1$ supercell of Zn-barlowite, and
  classical energies of a Heisenberg Hamiltonian with $J_1$ to $J_7$ as given in Table~\ref{tab:couplings} ($U=6$~eV line). }
\label{fig:magneticenergiesfit}
\end{figure}

Figures~\ref{fig:magneticconf} and \ref{fig:magneticenergiesfit} illustrate the energy mapping method used to extract Heisenberg Hamiltonian parameters from DFT total energies. Energy mapping is a widely used standard method in the context of spin systems~\cite{Rosner2002, Drechsler2007, HerbertsmithiteExchange, BarlowiteTheoryExperiment, Azurite, Glasbrenner2015}.

All 4096 spin configurations for the twelve inquivalent Cu spins of a $\sqrt{2}\times \sqrt{2}\times 1$ supercell of doped barlowite with $P\,1$ symmetry are classified with respect to their classical exchange energy. Representatives for the 25 distinct energy values are chosen (one example is shown in Figure~\ref{fig:magneticconf}), and total DFT energies are calculated (filled circles in Figure~\ref{fig:magneticenergiesfit}). Fitting these energies to classical exchange energies yields the seven exchange couplings shown in the third line of Table~\ref{tab:couplings}, with small statistical errors. The quality of the fit is illustrated by the classical exchange energies, shown in Figure~\ref{fig:magneticenergiesfit} as pentagons.

\vfill


\end{document}